\documentclass[prl,twocolumn,superscriptaddress,showpacs,nobalancelastpage]{revtex4}

\usepackage{graphicx}

\begin{document}

\title{Experimental signature of phonon-mediated spin relaxation}

\author{T. Meunier}
\affiliation{Kavli Institute of Nanoscience, Delft University of
Technology,\\
PO Box 5046, 2600 GA Delft, The Netherlands}
\author{ I. T. Vink}
\affiliation{Kavli Institute of Nanoscience, Delft University of
Technology,\\
PO Box 5046, 2600 GA Delft, The Netherlands}
\author{L. H. Willems van Beveren}
\affiliation{Kavli Institute of Nanoscience, Delft University of
Technology,\\
PO Box 5046, 2600 GA Delft, The Netherlands}
\author{K-J. Tielrooij}
\affiliation{Kavli Institute of Nanoscience, Delft University of
Technology,\\
PO Box 5046, 2600 GA Delft, The Netherlands}
\author{R. Hanson}
\affiliation{Kavli Institute of Nanoscience, Delft University of
Technology,\\
PO Box 5046, 2600 GA Delft, The Netherlands}
\author{F. H. L. Koppens}
\affiliation{Kavli Institute of Nanoscience, Delft University of
Technology,\\
PO Box 5046, 2600 GA Delft, The Netherlands}
\author{H. P. Tranitz}
\affiliation{Institut f$\ddot{u}$r Angewandte und Experimentelle
Physik, Universit$\ddot{a}$t Regensburg, Regensburg, Germany}
\author{W. Wegscheider}
\affiliation{Institut f$\ddot{u}$r Angewandte und Experimentelle
Physik, Universit$\ddot{a}$t Regensburg, Regensburg, Germany}
\author{L. P. Kouwenhoven}
\affiliation{Kavli Institute of Nanoscience, Delft University of
Technology,\\
PO Box 5046, 2600 GA Delft, The Netherlands}
\author{L. M. K. Vandersypen}
\affiliation{Kavli Institute of Nanoscience, Delft University of
Technology,\\
PO Box 5046, 2600 GA Delft, The Netherlands}

\date{\today}

\begin{abstract}

We observe an experimental signature of the role of the phonons in
spin relaxation between triplet and singlet states in a two-electron
quantum dot. Using both the external magnetic field and the
electrostatic confinement potential, we change the singlet-triplet
energy splitting from 1.3~meV to zero and observe that the spin
relaxation time depends non-monotonously on the energy splitting. A
simple theoretical model is derived to capture the underlying
physical mechanism. The present experiment confirms that spin-flip
energy is dissipated in the phonon bath.

\end{abstract}


\maketitle

Relaxation properties of a quantum system are strongly affected by
the reservoir where energy is dissipated. This has been seen clearly
for electron spins embedded in nanostructures. Spin relaxation times
$T_1$ up to a few~ns have been observed for free electrons in a two
dimensional electron gas (2DEG) where energy is easily given to the
motion~\cite{OhnoPRL}. In quantum dots, the discrete orbital energy
level spectrum imposes other energy transfer mechanisms. Near zero
magnetic field, the electron spin can directly flip-flop with the
surrounding nuclear spins, inducing short $T_1$'s of the order of
$\mu$s~\cite{JohnsonNature}. When a small magnetic field is applied,
direct spin exchange with nuclei is suppressed. Lattice vibrations,
i.e. phonons, are expected to become the dominant reservoir in which
spin-flip energy can be dissipated. This dissipative mechanism is
inefficient and very long spin relaxation times
follow~\cite{FinleyNature, FujisawaNature, NatureReadout,
RonaldPRL,JohnsonNature}.

For spin relaxation involving phonons, two physical processes are
important. The coupling between electron orbitals and phonons is
responsible for dissipation of energy and the spin-orbit interaction
provides the essential mixing between different spin states so the
spin states are coupled through electron-phonon
interaction~\cite{KhaetskiiST,golovachlossPRL,BulaevLossPRB}. Energy
conservation requires that the phonon energy corresponds to the
energy separation between the excited and the ground spin state.
Changing the energy separation affects the efficiency of the
electron spin relaxation in two ways. First, since the phonon
density of states increases with energy, the relaxation rate is
expected to increase with energy as well. Furthermore, since the
electron-phonon interaction is highly dependent on the matching
between the size of the dot and the phonon
wavelength~\cite{BockelmannPRB,FujisawaScience}, we expect a
suppression of relaxation for very large and for very small phonon
wavelengths in comparison to the dot size. Mapping the relaxation
time $T_1$ as a function of the energy splitting between the two
spin states will provide insight in both the electron-phonon
interaction and the spin-orbit coupling as well as an understanding
of the limitations on $T_1$. This is of particular relevance in the
context of both spintronic and spin-based quantum information
processing devices \cite{LossDiVincenzo}.

\begin{figure}[!t]
\includegraphics[width=3.4in]{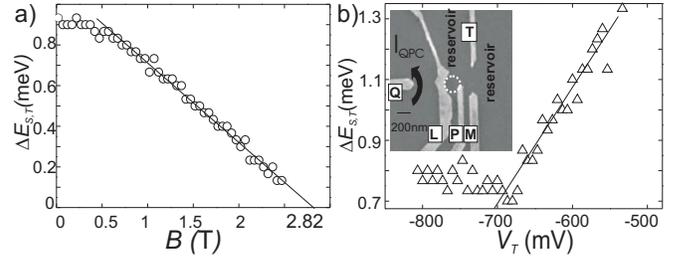}
\caption{a) Dependence of the energy splitting $\Delta E_{S,T}$  on
total magnetic field $B$. (b) Dependence of the energy splitting
$\Delta E_{S,T}$ on the voltage $V_T$ applied on gate 'T' at B$=0$.
Inset: Scanning electronic micrograph showing the sample design. The
2DEG, located 90~nm below the surface of a GaAs/AlGaAs
heterostructure, has an electron density of
$1.3\cdot10^{15}$~m$^{-2}$. By applying negative voltages to gates
$L$, $M$, $T$ and $Q$ we define a quantum dot (white dotted circle)
and a QPC. Gate $P$ is used to apply fast voltage pulses that
rapidly change the electrochemical potentials of the dot. We tune
the dot to the few-electron regime~\cite{CiorgaPRB}, and completely
pinch off the tunnel barrier between gates $L$ and $T$, so that it
is only coupled to one electron reservoir at a time~\cite{JeroAPL}.
A voltage bias of 0.7~mV induces a current through the QPC,
$I_{QPC}$, of about 30~nA. Tunneling of an electron on or off the
dot gives steps in $I_{QPC}$ of 300 pA~\cite{LievenAPL,EnsslinAPL}.
The QPC measurement bandwidth is 100~kHz.} \label{Fig1}
\end{figure}

Here, we study the spin relaxation time from triplet to singlet
states for different energy separations in a single quantum dot
containing two electrons. Singlet and triplet states have
respectively two electrons in the lowest orbital and one electron
each in the lowest and in the first excited orbital. In the
experiment, the energy splitting $\Delta E_{S,T}$ between these
two-electron spin states could be tuned from 0.9~meV to zero with a
perpendicular magnetic field  and from 0.9~meV to 1.3~meV by
deforming the dot potential~\cite{LeoFewEl,SachradjaGaCo}.

\begin{figure}[!t]
\includegraphics[width=3.4in]{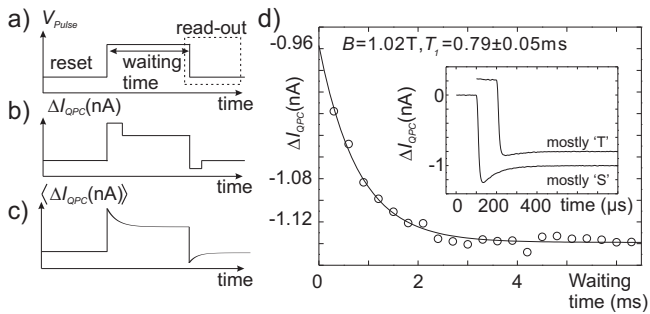}
\caption{(a) Voltage pulses applied to gate $'P'$ for the relaxation
measurement. The starting point is a dot with one electron in the
ground state (initialization). During the pulse, the singlet and
triplet electrochemical potentials are below the Fermi energy and a
second electron tunnels into the dot. Due to the difference in
tunnel rates~\cite{RonaldPRL}, most likely a triplet state will be
formed. We allow relaxation to occur during a waiting time that we
vary. After the pulse, both electrochemical potentials are moved
back above the Fermi energy and an electron tunnels out. This last
step allows us to read-out the spin state. (b) Schematic of the
$\Delta I_{QPC}$ induced by the voltage pulse on gate $'P'$. If the
state was singlet, a step from a slow tunneling event is added to
the QPC response just after the read-out pulse. If the state was
triplet, the tunneling event is too fast to be observed. (c) After
averaging over many single traces, a dip is observed and its
amplitude is proportional to the probability of having singlet
present in the dot. (d) Relaxation curve obtained for $B=1.02$~T
constructed by plotting the dip amplitude of the averaged traces at
a pre-defined time after the read-out pulse. The relaxation time,
$T_1=0.79\pm0.05$~ms, is extracted from an exponential fit to the
data (all the data are taken with a 100~kHz low-pass filter).
Inset : curve resulting from the averaging over 500
individual traces for the longest waiting time ($20$~ms) and for the
shortest waiting time ($300~\mu$s), offset by 100~$\mu$s and 0.2~nA
for clarity.} \label{Fig2}
\end{figure}

All the experiments are performed in a dilution refrigerator with a
quantum dot and a quantum point contact (QPC) defined in a 2DEG (see
inset of Fig.~\ref{Fig1}(b)). The conductance of the QPC is tuned to
about $e^2/h$, making it very sensitive to the charge on the
dot~\cite{FieldPRL}. The sample is mounted at an angle
$\phi=68^\circ \pm 5^\circ$ with respect to the direction of the magnetic field
$B$ where $\phi$ is derived from Shubnikov-deHaas oscillations. The
magnetic field component perpendicular to the 2DEG is equal to
$B\cos\phi$ ($\sim0.38B$). The electron temperature was measured to
be $180$~mK from the width of the Coulomb peaks. The lattice
temperature was $50$~mK.

We extract experimentally the energy splitting $\Delta E_{S,T}$
between the singlet and the triplet states as a function of both
magnetic field $B$ and the confinement potential using a pulse
spectroscopy technique \cite{JeroAPL}. The dependence of $\Delta
E_{S,T}$ on $B$ is presented in Fig.~\ref{Fig1}(a). Up to $0.4$~T,
$\Delta E_{S,T}$ does not vary significantly with magnetic field
which we relate to the elliptic nature of the dot at zero magnetic
field~\cite{TaruchaFewelecton}. For $B$ larger than 0.4~T, $\Delta
E_{S,T}$ decreases, to a good approximation, linearly with magnetic
field. For energy separations below 100~$\mu$eV, the thermal broadening of the reservoir
prevents us to measure $\Delta E_{S,T}$. From extrapolation of the
data, we can determine the magnetic field needed for singlet and
triplet energy levels to cross: $2.82\pm0.07$~T.

\begin{figure}[!t]
\includegraphics[width=3in]{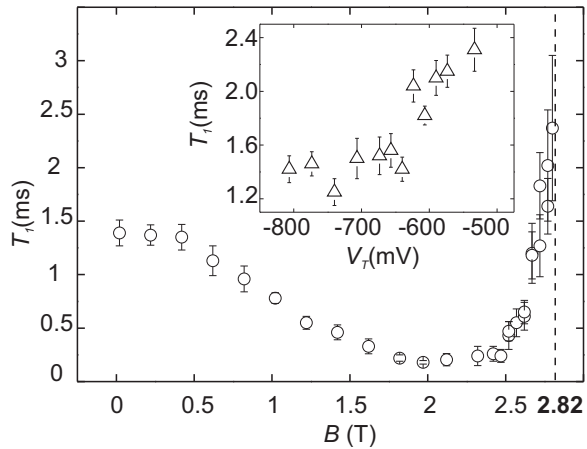}
\caption{The spin relaxation time $T_1$ as a function of the total
magnetic field. The magnetic field where singlet and triplet states
are degenerate is indicated by the dashed line. A minimum in $T_1$
is observed around 2.2~T. The error bars represent 70\% confidence
intervals. For energy separations close to degeneracy, the
sensitivity of the measurement is reduced and the uncertainty in
$T_1$ increases. Inset: dependence of the relaxation time $T_1$ on
$V_T$ at $B=0$~T.} \label{Fig3}
\end{figure}

We measure the relaxation time for varying $\Delta E_{S,T}$. To be
able to measure $T_1$ close to the degeneracy point, we use a
tunnel-rate selective read-out procedure (TRRO) \cite{RonaldPRL}
(see Fig.~\ref{Fig2}). The measured spin relaxation time $T_1$ as a
function of $B$ is presented in Fig.~\ref{Fig3}. The shape of the
$T_1$ dependence on magnetic field exhibits a striking
non-monotonous behavior. From $0.4$~T to $\sim 2$~T, corresponding
to a decrease in the energy splitting from 0.8~meV to 0.2~meV, the
relaxation time first decreases, reaching a minimum of $180~\mu$s.
In between 2~T and the degeneracy point (2.82~T), $T_1$ increases
whereas the energy splitting continues to decrease.

As a complementary study, we change $\Delta E_{S,T}$ in a different
way by controlling the electrostatic potential of the dot via the
voltage $V_T$ applied to gate 'T' and again look at $T_1$. The
dependence of $\Delta E_{S,T}$ on $V_T$ is presented in
Fig.~\ref{Fig1}(b). With this second experimental knob, $\Delta
E_{S,T}$ can be varied from 0.8~meV to 1.3~meV. We interpret the
change in the observed energy splitting as a consequence of a change
in the dot ellipticity. A more positive $V_T$ implies a more
circular dot and a larger energy splitting. We observe that $T_1$
further increases with $\Delta E_{S,T}$ as $V_T$ is varied at
$B=0$~T (see the inset of Fig.~\ref{Fig3}). The maximum energy
splitting reached at -530~mV, 1.3~meV, corresponds to a maximum of
$T_1=2.3$~ms. With both experimental knobs, we observe that when
$\Delta E_{S,T}$ is constant, $T_1$ is constant too (respectively
for $V_T<-650~$mV and $B<0.4~$T). These observations clearly
indicate that the most important parameter for the variation in the
triplet-singlet relaxation time is their energy separation.

The observed minimum in $T_1$ is precisely what one would expect for
energy relaxation mediated by the electron-phonon
interaction~\cite{BockelmannPRB,golovachlossPRL}. Indeed, the energy
splitting $\Delta E_{S,T}$ determines the relevant acoustic phonon
energy (acoustic phonons are the only available phonons for the
explored energy range). At $B$ $\sim2$~T, $\Delta
E_{S,T}\sim$~0.3~meV, the associated half-wavelength, approximately
30~nm (the group velocity for acoustic phonons $c_s\sim~$4000~m/s),
is comparable to the expected size of the dot and therefore the
coupling of the electrons in the dot to phonons is strongest. For
energy separations smaller (larger) than 0.3~meV, the phonon
wavelength is larger (smaller) than the size of the dot, the
coupling to the orbitals becomes smaller and $T_1$ increases. Taken
together, all these observations strongly suggest that the phonon
bath is the dominant reservoir for dissipating the spin-flip energy
during relaxation.

In order to get more insight in the role of the phonon wavelength,
we present a simplified model of the energy relaxation process
between triplet and singlet as a function of their energy splitting
$\Delta E_{S,T}$. From Fermi's golden rule, the relaxation rate
between the triplet and the singlet states with energy separation
$\Delta E_{S,T}$ is proportional to their coupling strength through
electron-phonon interaction and to the phonon density of states at
the energy $\Delta E_{S,T}$~\cite{BockelmannPRB}. To obtain a simple
analytical expression, we assume that the only effect of the
perpendicular magnetic field, the Coulomb interaction between
electrons and the modification of the potential landscape is to
change the energy splitting. Especially, their effects on the
spatial distribution of the wavefunctions are neglected and we
neglect the Zeeman energy. Furthermore, we restrict the state space of
the analysis to $|T_-\rangle$, $|T_+\rangle$, $|T_0\rangle$ and
$|S\rangle$ constructed from the lowest energy orbital and the first
excited orbital (even though the contributions to triplet-singlet
relaxation from higher orbitals can in fact be important
\cite{Losscitation}). In the notation $|T_-\rangle$, $|T_+\rangle$,
$|T_0\rangle$ and $|S\rangle$, both the orbital part
(assuming Fock-Darwin states) and the spin
part are present. Finally, we also neglect higher order (e.g.
two-phonon) processes, which are important at small magnetic field
\cite{ShonPRL}.

In contrast to the one electron
case~\cite{BulaevLossPRB,KhaetskiiST}, the spin-orbit interaction
admixes directly the first excited states $|T_\pm\rangle$ with the
ground state $|S\rangle$. Due to the selection rules of the
spin-orbit interaction, it does not affect $|T_0\rangle$ in lowest
order \cite{HawrylakPRB}. As a consequence, the spin relaxation time of
$|T_0\rangle$ can be much longer than $|T_\pm\rangle$
\cite{FujisawaPRL}. However, we do not observe any signature of a
slowly relaxing component in the experiment~\cite{note1}. Since the
spin-orbit coupling strength $M_{SO}$ is small in comparison with $\Delta
E_{S,T}$ (in the range accessed in the experiment), we can
approximate the new eigenstates of the system as:

\begin{displaymath}
\begin{array}{lll}
|S'\rangle=|S\rangle-\frac{M_{SO}}{\Delta
E_{S,T}}(|T_+\rangle+|T_-\rangle)
\\ \\
|T'_\pm\rangle= |T_\pm\rangle+\frac{M_{SO}}{\Delta E_{S,T}}|S\rangle
\end{array}
\end{displaymath}

\begin{figure}[!t]
\includegraphics[width=3in]{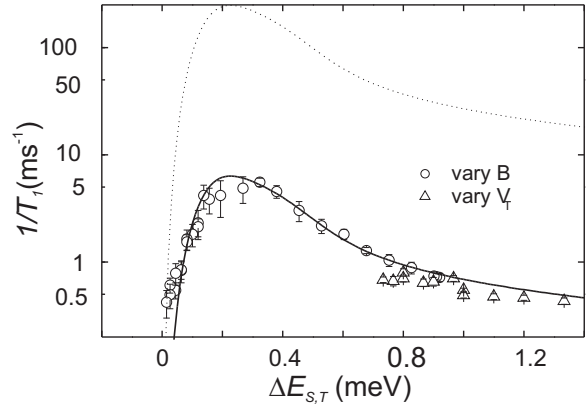}
\caption{Relaxation rate as a function of the energy splitting
$\Delta E_{S,T}$ deduced from the experimental data. The circles and
the triangles correspond to the experiment where we vary
respectively the magnetic field and the dot potential. The solid
(dotted) line is the curve for $M_{SO}=0.37~\mu$eV
($M_{SO}=2.31~\mu$eV) obtained from the simplified model.}
\label{Fig4}
\end{figure}

In general, $M_{SO}$ is dependent on the magnetic field, but to
simplify the discussion, we neglect this dependence
\cite{Losscitation,ShonPRL}. Since the electron-phonon interaction
preserves the spin, the coupling between $|T'_\pm\rangle$ and
$|S'\rangle$ has the following form:

\begin{displaymath}
\langle T'_\pm |H_{e,p}|S'\rangle= \frac{M_{SO}}{\Delta
E_{S,T}}(\langle S|H_{e,p}|S\rangle-\langle
T_\pm|H_{e,p}|T_\pm\rangle)
\end{displaymath}
where $H_{e,p}\sim
e^{i\mathbf{q}.\mathbf{r_1}}+e^{i\mathbf{q}.\mathbf{r_2}}$ is the
interaction Hamiltonian between electrons and phonons, $\mathbf{q}$
the phonon wavevector and $\mathbf{r_i}$ the positions of the
electrons. One can then interpret the coupling between
$|T'_\pm\rangle$ and $|S'\rangle$ as the difference of the
electron-phonon interaction strength for the corresponding
unperturbed states $|T_\pm\rangle$ and $|S\rangle$. If the phonon
wavelength is much larger than the dot size, the coupling to the
phonons is the same for both states and the two terms will cancel.
If the phonon wavelength is much shorter than the dot size, the
coupling is small for each state separately.

To provide a quantitative comparison to the data, we need to model
the electron-phonon interaction. Following
\cite{BockelmannPRB,golovachlossPRL}, we assume bulk-like 3D
phonons. For the energy separations discussed in our experiment,
only acoustic phonons are relevant. The Hamiltonian $H_{e,p}$ has
then the following expression:

\begin{displaymath}
H_{e,p}=\sum_{j,\mathbf{q}}\frac{F_z(q_z)}{\sqrt{2\rho q
c_j/\hbar}}(e^{iq_\parallel r_1}+e^{iq_\parallel
r_2})(e\beta_{j,\mathbf{q}}-iq\Xi_{j,\mathbf{q}})
\end{displaymath}
where $(\mathbf{q},j)$ denotes an acoustic phonon with wave vector
$\mathbf{q}=(\mathbf{q_\parallel},q_z)$, $j$ the phonon branch index
and $\rho=5300$~kg/m$^{3}$ is the density of lattice atoms. The
factor $F_z(q_z)$ depends on the quantum well geometry and is
assumed to be 1 in our model \cite{BockelmannPRB}. The speed of
sound for longitudinal and transverse phonons are respectively
$c_l=4730$~m/s and $c_t=3350$~m/s. We consider both piezo-electric
and deformation potential types of electron-phonon interaction. In
the considered crystal, the deformation potential interaction is
non-zero only for longitudinal phonons (with a coupling strength
$\Xi=6.7$~eV). In contrast, all phonon polarizations $j$ are
important for piezo-electric coupling. The coupling strength depends
on $\theta$, defined as the angle between the wavevector and the
growth axis, and varies for different polarizations as
$e\beta_{j,\mathbf{q}}=A_j(\theta) e\beta$ where
$e\beta=1.4\times10^9~$eV/m \cite{golovachlossPRL,note3}. Due to the
different dependence on $q$ for both mechanisms ($\sqrt{q}$ for
deformation potential interaction, $1/\sqrt{q}$ for piezo-electric
interaction), the piezo-electric (the deformation potential)
coupling between electrons and phonons is dominant for energy
separations below (above) $0.6$~meV. From direct application of
Fermi's golden rule, we derive the following analytical expression
for the spin relaxation rate $1/T_1$ :

\begin{displaymath}
\begin{array}{lll}
1/T_1=\frac{M_{SO}^2\alpha^4}{32\pi\rho\hbar^6
}\Bigl(\frac{\Xi^2\Delta
E_{S,T}^5}{\hbar^2c^9_{l}}\int^{\pi/2}_{0}d\theta \sin^5\theta
~e^{-\frac{\Delta
E_{S,T}^2\alpha^2\sin^2\theta}{2\hbar^2c_l^2}} \\
+\sum_j\frac{e^2\beta^2\Delta
E_{S,T}^3}{c^7_{j}}\int^{\pi/2}_{0}d\theta
|A_j(\theta)|^2\sin^5\theta~e^{-\frac{\Delta
E_{S,T}^2\alpha^2\sin^2\theta}{2\hbar^2c_j^2}}\Bigr)
\end{array}
\end{displaymath}
where $\alpha$ is the dot radius (in our model $\alpha$ is
independent of $\Delta E_{S,T}$ and is estimated to be $23$nm, from
the measured single particle level spacing). This simple model
reproduces the most important feature in the measurements, which is
that the coupling to the phonons vanishes for large and small energy
separations and is strongest when the phonon wavelength matches the
dot size (see Fig.~\ref{Fig4}).

The spin-orbit strength $M_{SO}$ appears in the expression of
$1/T_1$ only as a scaling factor. With a value $M_{SO}=0.4~\mu$eV
(corresponding to a spin-orbit length equal to $\hbar/2\alpha
m^*M_{SO}\approx50~\mu$m), the model reproduces the peak amplitude
of the data quite well (Fig.~\ref{Fig4}, solid line). However, this
value for $M_{SO}$ is about six times smaller than the values
reported in \cite{MarcusSO,KastnerSO} (the dotted line in
Fig.~\ref{Fig4} corresponds to the relaxation rate using this value
of $M_{SO}$ in the model). The discrepancy could be the result of
the exclusion of higher orbitals and the magnetic field dependence
of $M_{SO}$ in our model \cite{Losscitation,ShonPRL}. Again, we
emphasize that both curves have a maximum corresponding to a phonon
wavelength matching the dot size.

For single electron spin states, comparable variations of $T_1$ with
the energy splitting are expected although direct spin-orbit
coupling between Zeeman sublevels of the same orbital is zero. To
maximize the relaxation time of electron spin qubits, one needs then
to choose an energy separation between the spin states such that the
corresponding phonon wavelength is different from the dot size. To
complete our study of spin relaxation, it will be interesting to
rotate the sample with respect to the magnetic field since the spin-orbit
coupling strength depends on the angle between the crystallographic
axis and the magnetic field~\cite{KhaetskiiST,
golovachlossPRL,AltshulerPRL}.

\begin{acknowledgments}

We thank V. Golovach and D. Loss for drawing our attention
to the role of the phonon wavelength in spin relaxation and for
useful discussions; R. Schouten, B. van der Enden and W. den
Braver for technical assistance. Supported by the Dutch Organization
for Fundamental Research on Matter (FOM), the Netherlands
Organization for Scientific Research (NWO), the DARPA QUIST program
and a E.U. Marie-Curie fellowship (T.M.).

\end{acknowledgments}


\begin{references}
\bibitem{note1}
By applying the complete relaxation measurement protocol presented
in Fig.~\ref{Fig2}(a,b), we observe experimentally a reduction of
the read-out visibility (the contrast between the signal for short
and long waiting time) when $\Delta E_{S,T}$ is varied. Several
reasons can be considered to explain the reduction of visibility
observed in the experiment. At zero magnetic field, the difference
between the singlet and triplet tunnel rates can be increased as the
dot is somewhat elliptical. At relatively high magnetic fields, the
dot loses its elliptical nature and then the measurement loses
visibility. Another possible explanation is related to the slow
spin-orbit decay from $|T_0\rangle$ to $|S\rangle$ in comparison
with the longest waiting time we explored (20~ms). Therefore, the
$T_0$ population does not contribute to the visibility unless
$|T_0\rangle$ gets first mixed with $|T_\pm\rangle$. As $B$
increases the mixing between triplets gets suppressed and the
visibility goes down.
\end{references}



\end{document}